\documentclass[aps,prd,floatfix,superscriptaddress,preprintnumbers]{revtex4}
\usepackage{amssymb}
\usepackage{amsmath}
\usepackage{amsfonts}
\usepackage{epsfig}             
\usepackage{graphicx}
\usepackage{tabularx}
\usepackage{bm}
\usepackage{color}
\newcommand{\seq}{\begin{subequations}}
\newcommand{\sen}{\end{subequations}}
\newcommand{\eq}{\begin{eqnarray}}
\newcommand{\en}{\end{eqnarray}}

\begin{document}

\title{Bulk-to-boundary propagators with arbitrary \\
  total angular momentum $J$ in soft-wall AdS/QCD} 

\author{Valery E. Lyubovitskij}
\affiliation{Institut f\"ur Theoretische Physik,
Universit\"at T\"ubingen,
Kepler Center for Astro and Particle Physics,
Auf der Morgenstelle 14, D-72076 T\"ubingen, Germany}
\affiliation{Departamento de F\'\i sica y Centro Cient\'\i fico
Tecnol\'ogico de Valpara\'\i so-CCTVal, Universidad T\'ecnica
Federico Santa Mar\'\i a, Casilla 110-V, Valpara\'\i so, Chile}
\affiliation{Millennium Institute for Subatomic Physics at
the High-Energy Frontier (SAPHIR) of ANID, \\
Fern\'andez Concha 700, Santiago, Chile}
\author{Ivan Schmidt}
\affiliation{Departamento de F\'\i sica y Centro Cient\'\i fico
Tecnol\'ogico de Valpara\'\i so-CCTVal, Universidad T\'ecnica
Federico Santa Mar\'\i a, Casilla 110-V, Valpara\'\i so, Chile}

\begin{abstract}

We derive the equations of motion for the bulk-to-boundary propagators
of the anti-de Sitter (AdS) boson and fermion fields with arbitrary
total angular momentum $J$, in a soft-wall AdS/QCD model and solve it
analytically. It provides the opportunity to study transition form
factors induced by these bulk-to-boundary propagators, both for on-shell
and off-shell hadrons. This is a continuation of our study of hadron form
factors induced by the bulk-to-boundary propagator with total angular
momentum $J=1$ (e.g., electromagnetic form factors of mesons, nucleons,
and nucleon resonances). 

\end{abstract}

\maketitle

\section{Introduction}

The soft-wall anti-de Sitter/quantum chromodynamics (AdS/QCD) model
proposed in Ref.~\cite{Karch:2006pv}
plays an important role for the description and understanding of
hadron structure: mass spectra, form factors, parton distributions, 
QCD scattering processes (like Drell-Yan, deep-inelastic scattering),
etc. The pioneer contribution to the investigation of QCD scattering
processes based on gauge/string duality was made
in Ref.~\cite{Polchinski:2002jw}, and the
success of the soft-wall model is based on the fact that 
it provides analytical calculations 
of hadronic properties. The formalism of the soft-wall
AdS/QCD model is based on phenomenological actions 
formulated in terms of boson and fermion AdS fields, propagating
in five-dimensional AdS space. One should stress that the 
Hamiltonian approach is also widely used, especially in connection with
the light-cone formalism, e.g., in the model of Ref.~\cite{Brodsky:2007hb}. 
Four of the five dimensions of
the AdS space correspond to the Minkowski subspace
and the fifth (holographic) dimension $z$ corresponds to a scale.  
Conformal and chiral symmetry in the underlying actions are broken
by introducing the dilaton field, quadratically dependent on the variable
$z$ in the exponential prefactor of the action or in the
phenomenological potential. In the case of the Hamiltonian
approach~\cite{Brodsky:2007hb} the conformal symmetry 
of the Hamiltonian remains.
Based on this action one can solve
two problems: (i) The bound-state problem, i.e. 
derive equations of motion (EOM) for the bulk profiles $\phi(z)$
(the parts of the AdS fields explicitly dependent 
on the holographic variable $z$).  
These profiles obey Schr\"odinger type equations of motion, which
are solved analytically~\cite{Karch:2006pv}. The solutions of these
equations correspond to the hadronic mass spectrum due to the duality
of bulk profiles and hadronic wave function. 
(ii) The scattering problem, i.e. 
one can derive EOM for the bulk-to-boundary propagators $V(-q^2,z)$
describing the momentum dependence of the AdS field traveling
from the AdS interior to its boundary (Minkowski space).
In particular, the bulk-to-boundary
propagators depend on two variables: holographic coordinate $z$ 
and $q$, which is Fourier conjugate to Minkowski coordinate $x$.
Therefore, the main components produced by
the AdS/QCD soft-wall action needed for the study of hadron structure
are the bulk profiles $\phi(z)$, dual to hadronic wave functions
describing the hadrons on the mass shell,  
and the bulk-to-boundary $V(-q^2,z)$, dual to the off-shell
external gauge fields or external hadrons. In particular,
hadronic form factors, which are the main focus of the present paper,
are the integrals over $z$ of the product of bulk-to-boundary
propagators and two bulk profiles. One should stress that
up to now, the study of hadronic form factors has been focused on
the quantities induced by the bulk-to-boundary propagator with 
the total angular momentum  $J=1$, dual to electromagnetic field. 
In particular, following the idea proposed in Ref.~\cite{Erlich:2005qh}
one can derive the massless bulk-to-boundary
propagator for the field with total angular momentum $J=1$,
and reach its application for the calculation of form factors of mesons
and baryons in Euclidean spacetime, which has been done very successfully
in Refs.~\cite{Brodsky:2007hb},\cite{Grigoryan:2007my}-\cite{Lyubovitskij:2020gjz}.  
For application of the bulk-to-boundary propagators with $J=0$ and
$J=1$ in Minkowski space see, e.g., Ref.~\cite{MartinContreras:2018nhv}. 
In fact, the soft-wall AdS/QCD not only provides the correct power scaling
description of form factors and helicity amplitudes of all hadrons
at large~$Q^2$ in Euclidean spacetime~\cite{Brodsky:1973kr};
it is also able to give good agreement with the data at low and
intermediate $Q^2$. Note that up to now only the massless AdS bulk-to-boundary
propagator, dual to massless gauge vector fields, has been considered
in the context of soft-wall AdS/QCD. 

The main objective of the present paper is to extend the soft-wall model
formalism for the study of bulk-to-boundary propagators 
with arbitrary $J$ in the Euclidean spacetime.
First, we consider massless bulk-to-boundry propagators, which
are relevant for the description of gravitons or light hadrons for which
one can apply the massless limit. Second, we extend our results for
the case of massive bulk-to-boundary propagators dual to massive
gauge bosons and massive hadrons. As a result, we derive analytical
expressions of the form factors describing:
(i) the coupling of off-shell massless gauge bosons (photon, graviton) or
massless scalar/pseudoscalar fields of new physics (NP):
axions, axionlike particles (ALPs),
etc. with two on-shell hadrons; (ii) the coupling of off-shell massive
gauge bosons (weak $W^\pm$ and $Z^0$ bosons) or Higgs $H$,
with two on-shell hadrons;
(iii) the coupling of off-shell massless hadrons with two on-shell hadrons;
(iv) the coupling of off-shell massive hadrons with two on-shell hadrons.
In all cases the off-shell behavior of gauge fields and hadrons is encoded
in the $Q^2$ behavior of the corresponding bulk-to-boundary propagator.
This provides an opportunity to study the off-shell behavior
of hadronic form factors, i. e. direct coupling of three particles,
when one particle is off-shell and the other two are on-shell.
It provides useful insight to lattice QCD and effective field theories,
such as chiral perturbation theory (ChPT), heavy hadron ChPT,
where direct couplings of hadrons are calculated from first principles
(lattice QCD) or provide input parameters for phenomenological Lagrangians.

The paper is organized as follows.
In Sec. II, we discuss the derivation of bulk-to-boundary
propagators dual to off-shell gauge fields and hadrons.
First, we consider the case of boson propagators and then we extend
our formalism to the case of fermions. For massive gauge fields
and hadrons we propose an extension of the bulk-to-boundary
propagators to a massive case. Finally, in Sec. III we present
our conclusion. 

\section{Formalism}

\subsection{Boson bulk-to-boundary propagator}

We start by specifying the AdS$_5$ metric 
\eq
ds^2 = 
g_{MN} dx^M dx^N = \eta_{ab} \, e^{2A(z)} \, dx^a dx^b = e^{2A(z)}
\, (\eta_{\mu\nu} dx^\mu dx^\nu - dz^2)\,, \hspace*{1cm}
\eta_{\mu\nu} = {\rm diag}(1, -1, \ldots, -1) \,,
\en
where $M$ and
$N = 0, 1, \ldots , 4$ are the base manifold indices,
$a=(\mu,z)$ and $b=(\nu,z)$ are the local Lorentz (tangent) indices,
and $g_{MN}$ and  $\eta_{ab}$ are curved and flat metric tensors,
which are related by the vielbein
$\epsilon_M^a(z)= e^{A(z)} \, \delta_M^a$ as
$g_{MN} =\epsilon_M^a \epsilon_N^b \eta_{ab}$.
Here $z$ is the holographic coordinate. 
We restrict our discussion to a conformal-invariant metric with $A(z) = \log(R/z)$,
where $R$ is the AdS radius. 

The action of the soft-wall AdS/QCD model describing totally
symmetric traceless bosonic fields $V_{M_1 \cdots M_J}(x,z)$
with arbitrary integer $J$, was derived in Ref.~\cite{Karch:2006pv}.
In particular, this action has a simplified form in the axial
gauge $V_{z \cdots}(x,z) = 0$: 
\eq\label{actionV}
S_J &=& \frac{(-)^J}{2}
\int d^4 x dz \ e^{- B_J(z)} \,  
\partial_MV_{\mu_1 \cdots \mu_J}(x,z) \,
\partial^MV^{\mu_1 \cdots \mu_J}(x,z) \,, 
\en
where $\partial_M \otimes \partial^M = \partial_\mu \otimes \partial^\mu
- \partial_z \otimes \partial_z$, 
$B_J(z) = \varphi(z) - (2J-1) A(z)$,
$\varphi(z) = \kappa^2 z^2$ is the dilaton field, 
and $\kappa \sim 500$ MeV~\cite{Brodsky:2007hb,Branz:2010ub,Gutsche:2012ez}
is the dilaton scale parameter. 

The massless bulk-to-boundary propagator $V_J(q,z)$ of the $V_{\mu_1 \cdots \mu_J}(x,z)$
field is given by the Fourier transformation:
\eq
V_{\mu_1 \cdots \mu_J}(x,z) = \int \frac{d^4 q}{(2\pi)^4}  \,
e^{-iqx} \, V_{\mu_1 \cdots \mu_J}(q)  \, V_J(-q^2,z)  \,, 
\en
where $q$ is the Fourier conjugates to $x$.  
Next, from Eq.~(\ref{actionV}) we derive the equation of motion
for the propagator $V_J(-q^2,z)$: 
\eq\label{bulkV}
\partial_z \biggl( e^{-B_J(z)}  \, \partial_z \, V_J(-q^2,z) \biggr)
+ e^{-B_J(z)}  \, q^2 \, V_J(-q^2,z) = 0 \,.
\en
which has an analytical solution in terms of gamma $\Gamma(a)$
and Trikomi $U(a,b,z)$ functions:
\eq\label{propagator_VJ}
V_J(Q^2,z) &=& (\kappa^2 z^2)^J \
\frac{\Gamma(a_J+1)}{\Gamma(J)} \ U(a_J+1,J+1,\kappa^2 z^2) \,,
\en
where $a_J = a + J-1$, $a = Q^2/(4 \kappa^2)$, $Q^2 = -q^2$ is the
Euclidean momentum squared, and an integral representation
for the Trikomi function reads
\eq
U(a,b,c) &=& \frac{1}{\Gamma(a)} \, \int\limits_0^\infty dt  \, e^{- c t} \,
t^{a-1} \, (1+t)^{b-a-1}
\nonumber\\
&=& \frac{1}{\Gamma(a)} \,
\int\limits_0^1  \frac{dx \, x^{a-1}}{(1-x)^b}
\, e^{- \frac{c x}{1-x}} \,. 
\en
Therefore, an integral representation for the propagator $V_J(Q^2,z)$
is given by
\eq\label{propagator_VJ_int}
V_J(Q^2,z)
= \frac{(\kappa^2 z^2)^J}{\Gamma(J)} \,
\int\limits_0^1  \frac{dx \, x^{a_J}}{(1-x)^{J+1}}  \,
e^{-\frac{\kappa^2 z^2 x}{1-x}} \,.
\en
By changing the integration variable $x=y/(y+\kappa^2 z^2)$
one can derive another representation for
the $V_J(Q^2,z)$
\eq\label{propagator_VJ_int2}
V_J(Q^2,z) &=& \frac{1}{\Gamma(J)} \,
\int\limits_0^\infty  dy \, y^{J-1} \, e^{-y} \,
\biggl(\frac{y}{y+\kappa^2 z^2}\biggr)^{a} 
\,. 
\en
Additional useful integral representation for the $V_J(Q^2,z)$,
derived from Eq.~(\ref{propagator_VJ_int}) by partial
integration, reads
\eq\label{propagator_VJ_int3}
V_J(Q^2,z) &=& \frac{1}{B(a,J)} \,
\int\limits_0^1  dx \, x^{a-1} \,
(1-x)^{J-1}  \,
e^{-\frac{\kappa^2 z^2 x}{1-x}} \,, 
\en
$B(x,y) = \Gamma(x) \Gamma(y)/\Gamma(x+y)$ is the beta function. 

Now let us consider the properties of the derived bulk-to-boundary
propagators.
A nice feature of the derived bulk-to-boundary propagator
$V_J(Q^2,z)$ is the following: while it was derived for the boson
propagators with higher $J \ge 2$, it is also valid
for the limit $J=1$. 
In particular, in the limit $J=1$ (the case of the
vector bulk-to-boundary propagator), $V_1(Q^2,z)$ reduces
to the well-known result~\cite{Grigoryan:2007my,Gutsche:2011vb}
\eq
V_1(Q^2,z) &=& \kappa^2 z^2 \, \Gamma(a+1) \, U(a+1,2,\kappa^2 z^2)
\nonumber\\
&=&  \kappa^2 z^2 \,
\int\limits_0^1  \frac{dx \, x^{a}}{(1-x)^2}  \, e^{-\frac{\kappa^2 z^2 x}{1-x}}
= \int\limits_0^\infty  dy \, e^{-y} \,
\biggl(\frac{y}{y+\kappa^2 z^2}\biggr)^{a} \,. 
\en
The vector bulk-to-boundary propagator obeys the important
conditions~\cite{Grigoryan:2007my,Gutsche:2011vb}:
(i) charge conservation $V_1(0,z) = 1$ at $Q^2=0$,
(ii) local limit $V_1(Q^2,0) = 1$ at $z=0$,
(iii) confinement $V_1(Q^2,z) \to 0$ at $z \to \infty$,
(iv) it produces power scaling of hadronic form factors
$F(Q^2) \sim 1/Q^{2 (\tau-1)}$ at large $Q^2$~\cite{Brodsky:1973kr},
where $\tau$ is the leading twist of the hadron,
which is also the number of its constituent partons. 

From the integral representation~(\ref{propagator_VJ_int2}), it immediately
follows that all properties (i)-(iv) relevant for the vector propagator
are also valid for the propagators with higher $J \ge 2$.   
In particular, from Eq.~(\ref{propagator_VJ_int2}) it follows
that the normalization conditions $V_J(Q^2,0) = V_J(0,z) = 1$ are independent
of $J$. Obviously, the propagator $V_J(Q^2,z)$ has no
proper limit $J=0$. In particular, while at $Q^2=0$ and $z=0$
the scalar propagator $V_0(Q^2,z)$ has the required  normalizations
$V_0(Q^2,0) = V_0(0,z) = 1$: it vanishes at finite values
of $Q^2$ and $z$. Therefore, we had to propose an action for
the scalar AdS field, which produces a scalar bulk-to-boundary
propagator consistent with the following requirements:
(i) normalization condition $V_0(Q^2,0) = V_0(0,z) = 1$, 
(ii) $V_0(Q^2,z)$ is finite at $Q^2 \neq 0$ and $z \neq 0$,
(iii) confinement $V_0(Q^2,z) \to 0$ at $z \to \infty$,
(iv) correct power scaling of hadronic form factors
$F(Q^2) \sim 1/Q^{2 (\tau-1)}$ at large $Q^2$~\cite{Brodsky:1973kr}. 
One of such actions, which obeys the above requirements,
reads 
\eq\label{scalar_action} 
S_0 &=& \frac{1}{2}
\int d^4 x dz \ e^{-B(z)} \,  
\partial_MS(x,z) \,
\partial^MS(x,z)
\,,  
\en
where $B(z) = \varphi(z) - A(z)$.

Next, from the action~(\ref{scalar_action}) we derive the
following equation of motion
for the scalar propagator $V_0(-q^2,z)$: 
\eq\label{bulkV0}
\partial_z \biggl( e^{-B(z)}  \, \partial_z \, V_0(-q^2,z) \biggr)
+ e^{-B(z)}  \, q^2 \, V_0(-q^2,z) = 0 \,,
\en
which has the solution  $V_0(Q^2,z)$, coinciding with the vector
bulk-to-boundary propagator $V_0(Q^2,z) \equiv V_1(Q^2,z)$. 

Note that at $Q^2 \to \infty$ the bulk-to-boundary propagator
$V_J(Q^2,z)$ for $J \ge 1$ behaves as
\eq\label{bulk_Q2_infinity}
V_J(Q^2,z)& \to & 
\frac{e^{\kappa^2 z^2}}{\Gamma(J)} \, 
\biggl(\frac{Q^2 z^2}{4}\biggr)^J \,
\int\limits_0^\infty  \frac{dt}{t^{J+1}} \,
\exp\biggl(- t - \frac{Q^2 z^2}{4t}\biggr)
\nonumber\\ 
&=&
\frac{2 e^{\kappa^2 z^2}}{\Gamma(J)} \, 
\biggl(\frac{Q z}{2}\biggr)^J \,
K_J(Q z)
\,,
\en
where $Q=\sqrt{Q^2}$, and
\eq
K_n(x) = \frac{x^n}{2^{n+1}} \,
\int\limits_0^\infty  \frac{dt}{t^{n+1}} \,
\exp\biggl(- t - \frac{x^2}{4 t}\biggr)   
\en
is the modified Bessel function of the second kind
for arbitrary $n$~\cite{Brodsky:2016uln}.
As before, the $Q^2 \to  \infty$ asymptotics
coincides for $J=0$ and $J=1$. 
It was shown in Ref.~\cite{Brodsky:2016uln} that
in the case $J=1$ and in the limit $\kappa \to 0$,
the vector bulk-to-boundary propagator $V_1(Q^2,z)$
in the soft-wall AdS/QCD model, 
\eq
V_1(Q^2,z) = Q z \, K_1(Q z)  \,, 
\en 
coincides with the one obtained in
the hard-wall AdS/QCD model~\cite{Erlich:2005qh}.
Therefore, we make the prediction that 
the $Q^2 \to  \infty$ asymptotics of the bulk-to-boundary
propagator in the hard-wall model for arbitrary $J \ge 1$
coincides with the one in the soft-wall model for $\kappa \to 0$: 
\eq
V_J(Q^2,z) = \frac{2}{\Gamma(J)} \, 
\biggl(\frac{Q z}{2}\biggr)^J \, K_J(Q z)
\,. 
\en 

We should stress that the massless boson bulk-to-boundary
propagator $V_J(Q^2,z)$ is mostly relevant for
the description of the propagation of massless
gauge bosons (photon with $J=1$ and graviton with $J=2$) and 
massless scalar/pseudoscalar of NP (axion, ALPs)
with $J=0$. In the case of massive gauge fields, such as the 
weak $W^\pm$ and $Z^0$ bosons or the Higgs $H$,
one should include their
masses, which appear after spontaneous breaking of gauge symmetry.
We propose to include the finite mass for the bulk-to-boundary
propagator, by shifting the square of the momentum as
$- q^2 = Q^2 \to - q^2 + M^2 = Q^2 + M^2$, where $M$ 
is the mass of the gauge field or Higgs, taken from data in Ref.~\cite{PDG22}:
\eq
M_{W^\pm} = 80.377 \pm 0.012 \ \mathrm{GeV}\,, \quad 
M_{Z^0}  = 91.1876 \pm 0.0021 \ \mathrm{GeV}\,, \quad
M_{H}  = 125.25 \pm 0.17 \ \mathrm{GeV}\,.
\en 
For example, the massive bulk-to-boundary propagator of the weak bosons
and Higgs reads
\eq\label{V_M2} 
V(Q^2+M^2,z) = \int\limits_0^\infty  dy \, e^{-y} \,
\biggl(\frac{y}{y+\kappa^2 z^2}\biggr)^{a(M^2)} 
\,, 
\en 
where $a(M^2) = a + M^2/(4 \kappa^2) = (Q^2 + M^2)/(4 \kappa^2)$
and $M = M_{W^\pm}, M_{Z^0}, M_{H}$.
One can see that our extension to massive bulk-to-boundary propagators is
consistent.
It is clear that for $M^2 \gg Q^2$ one can neglect 
the $Q^2$ dependence of the propagator $V(Q^2+M^2,z)$, i.e.
in this limit $V(Q^2+M^2,z) \to V(M^2,z)$ in consistency with the Standard Model (SM). 
Also, the massless limit $M \to 0$ is straightforward
leading to the massless propagator. Here we consider as example the particles
of the SM, but this discussion is true for any other
massless/massive gauge fields or other structureless particles
(axion, ALPs, etc.).  

Next we clarify how to use the bulk-to-boundary propagator with arbitrary $J$
for the description of the propagation of composite particles --- hadrons.
The difference of hadrons from structureless particles is that hadrons
in the soft-wall AdS/QCD are described by hadronic wave functions.
The latter are dual to the profiles of the AdS fields depending
on the holographic variable $z$. In particular, the meson wave
function describing the hadron, which is made by the Fock state with
leading twist $\tau$, reads~\cite{Brodsky:2007hb,Gutsche:2011vb,Branz:2010ub}  
\eq\label{Htaunwf}
\phi_{M_\tau}(z) = \sqrt{\frac{2}{\Gamma(\tau-1)}}
\, \kappa^{\tau-1} \, z^{\tau-3/2} 
e^{-\kappa^2z^2/2} \,. 
\en
We should stress that the inclusion of subleading Fock states
in the context of the soft-wall model has been considered
in Refs.~\cite{Brodsky:2011xx,Gutsche:2012bp,%
Gutsche:2015xva,Sufian:2016hwn,Lyubovitskij:2020gjz}. 
In the present paper we restrict the discussion to the leading Fock state
contribution to the structure of the specific hadron. 
Therefore, the massive bulk-to-boundary propagator with
total angular momentum $J$, dual to massive meson with the same $J$
and made by the leading Fock state with twist $\tau$,  
must be constructed as a product of the massive bulk-to-boundary
propagator 
\eq
V_J(Q^2+M_M^2,z) = 
\int\limits_0^\infty  dy \, y^{J-1} \, e^{-y} \,
\biggl(\frac{y}{y+\kappa^2 z^2}\biggr)^{a(M_M^2)} 
\en
where $M_M$ is the mass of a meson 
and meson wave function $\phi_{M_\tau}(z)$~(\ref{Htaunwf}).
For convenience, we denote the bulk-to-boundary propagator
dual to massive meson with arbitrary $J$ as
\eq\label{V_hadron}
\phi_{M_{J, \tau}}(Q^2+M_M^2,z) = V_J(Q^2+M_M^2,z) \, \phi_{M_\tau}(z) \,.
\en 
One can see that the massive bulk-to-boundary propagator, dual
to a hadron~(\ref{V_hadron}), obeys important requirements:  
(i)~mass-shell limit $q^2 = - Q^2 = M_M^2$ and (ii)~massless
limit $M_M^2 = 0$. In particular, in the limit (i):
$V_J(Q^2+M_M^2,z)  \to V_J(0,z) = 1$ and therefore
\eq
\phi_{M_{J, \tau}}(Q^2+M_M^2,z) \to \phi_{M_{J, \tau}}(0,z)
= \phi_{M_\tau}(z) \,. 
\en 
In the limit (ii): $V_J(Q^2+M_M^2,z)  \to V_J(Q^2,z)$
and therefore
\eq\label{phi_M_Jtau} 
\phi_{M_{J, \tau}}(Q^2+M_M^2,z) \to \phi_{M_{J, \tau}}(Q^2,z)
= V_J(Q^2,z) \, \phi_{M_\tau}(z)
\en

We summarize the results of this subsection. We derived the
set of massless and massive
bulk-to-boundary propagators with arbitrary $J$,
dual to massless and massive fields of SM (photon, weak bosons, Higgs)
and of NP (axion, ALPs, etc.) and of hadrons.
These quantities describe off-shell behavior of SM or NP particles and
of hadrons and could be used for the calculation of transition form factors
involving off-shell and on-shell states. As we stressed before, it could  
be useful to provide insight to lattice QCD and effective field theories,
where direct couplings of hadrons are calculated from first
principles or provide input parameters for phenomenological Lagrangians.
Later, we will extend our ideas to the sector of
fermion bulk-to-boundary propagators.

\begin{figure}
\begin{center}
\epsfig{figure=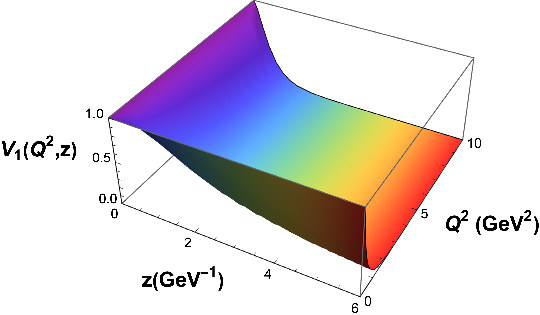,scale=.8}
\hspace*{.5cm}
\epsfig{figure=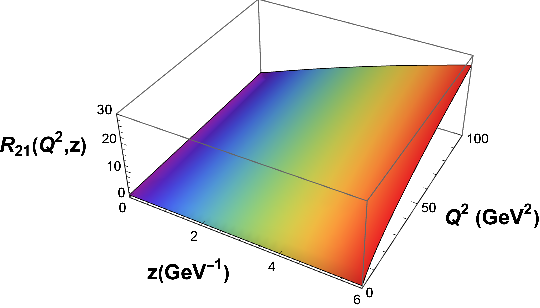,scale=.8} 
\end{center}
\vspace*{-.5cm}
\noindent
\caption{3D plots of the massless boson bulk-to-boundary propagators 
  as functions of $Q^2$ and $z$, which are dual to massless
  gauge fields with $J=1$ and $J=2$: 
  $V_1(Q^2,z)$ (left-upper panel), 
  ratio $R_{21}(Q^2,z) = V_2(Q^2,z)/V_1(Q^2,z)$  (right-upper panel). 
  \label{fig1}}

\begin{center}
\epsfig{figure=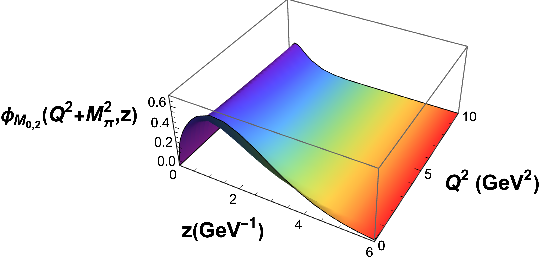,scale=.9}
\hspace*{.5cm}
\epsfig{figure=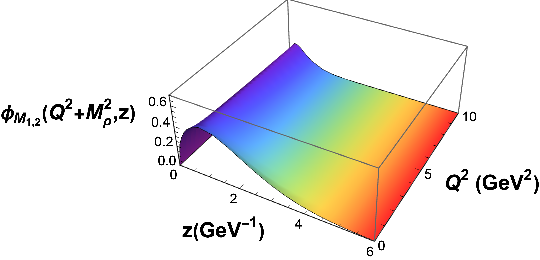,scale=.9}\\
\epsfig{figure=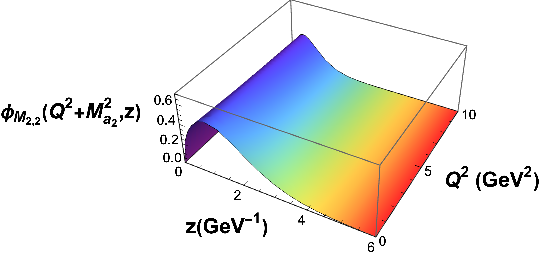,scale=.9}
\hspace*{.5cm}
\epsfig{figure=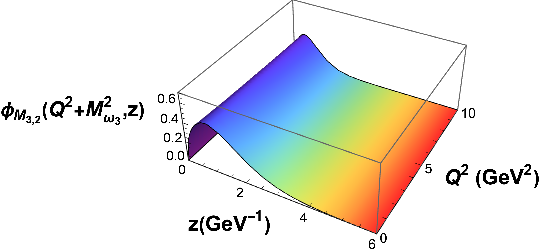,scale=.9}
\end{center}
\vspace*{-.5cm} 
\noindent
\caption{3D plots of the massive bulk-to-boundary
propagators $\phi_{M_{J, \tau}}(Q^2+M_M^2,z)$ as functions of $Q^2$ and $z$,
dual to massive mesons at fixed value of
the leading twist $\tau = 2$ for specific mesons: 
(a) pion with $J=0$ and $M_\pi = 0.13957$ GeV (left-upper panel),  
(b) $\rho$ meson with $J=1$ and $M_\rho = 0.7665$ GeV (right-upper panel), 
(c) $a_2$ meson with $J=2$ and $M_{a_2} = 1.3186$ GeV  (left-bottom panel), 
(d) $\omega_3$ meson with $J=3$ and $M_{\omega_3} = 1.67$ GeV
(right-bottom panel). 
\label{fig2}}

\begin{center}
\epsfig{figure=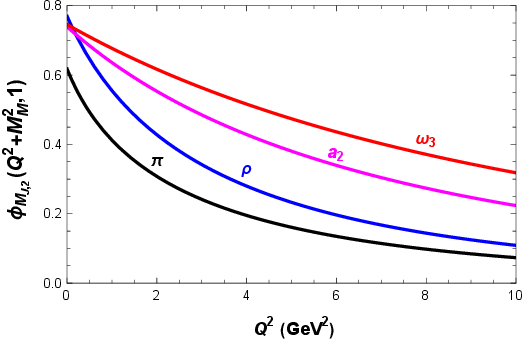,scale=.9}
\hspace*{1.5cm}
\epsfig{figure=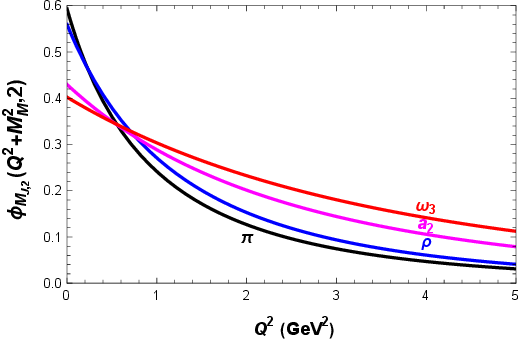,scale=.9}
\end{center}
\vspace*{-.5cm} 
\noindent
\caption{2D plots of the massive bulk-to-boundary 
propagators $\phi_{M_{J, \tau}}(Q^2+M_M^2,z)$  
as functions of $Q^2$ and $J=0,1,2,3$, 
dual to massive mesons at fixed values of 
the leading twist $\tau = 2$ and holographic coordinate: 
$z=1$ GeV$^{-1}$ (left panel), 
$z=2$ GeV$^{-1}$ (right panel). 
\label{fig3}}
\end{figure}

In Figs.~\ref{fig1}-\ref{fig3} we present two- and three-dimensional plots
illustrating the properties of boson bulk-to-boundary propagators, dual
to the gauge bosons with $J=1$ and $J=2$, and hadrons (mesons)
with arbitrary integer $J$. 
In particular, in Fig.~\ref{fig1} (left panel) we show results for the
bulk-to-boundary propagator $V_1(Q^2,z)$ as a function of $Q^2$ and $z$,
dual to the massless gauge field with $J=1$ (like photon).
In Fig.~\ref{fig1} (right panel) we show results for the ratio 
$R_{21}(Q^2,z) = V_2(Q^2,z)/V_1(Q^2,z)$ of two propagators with
$J=2$ and $J=1$, dual to the massless gauge fields with $J=2$ (graviton)
and $J=1$ (photon). One can see that the plot for $V_1(Q^2,z)$
decreases when $z$ and $Q^2$ increase. The ratio $R_{21}(Q^2,z)$
increases when $z$ and $Q^2$ increase. One should stress that
for both plots the $Q^2 \to \infty$ asymptotics is fully 
consistent with our analytical prediction in Eq.~(\ref{bulk_Q2_infinity}).
In particular, at large $Q^2 \to \infty$ the $R_{21}(Q^2,z)$ behaves as
\eq 
R_{21}(Q^2,z) = \frac{Q z}{2} \, \frac{K_2(Q z)}{K_1(Q z)} 
= \frac{Q^2 z^2}{4} \,
\frac{\int\limits_0^\infty  \dfrac{dt}{t^3} \,
\exp\Big(- t - \frac{Q^2 z^2}{4 t}\Big)}
{\int\limits_0^\infty  \dfrac{dt}{t^2} \,
\exp\Big(- t - \frac{Q^2 z^2}{4 t}\Big)} \,. 
\en 

In Fig.~\ref{fig2} we show results for the massive bulk-to-boundary
propagators $\phi_{M_{J, \tau}}(Q^2+M_M^2,z)$ as functions of $Q^2$ and $z$ 
dual to massive mesons at fixed value of
leading twist $\tau = 2$ for specific mesons
characterized by total angular momentum $J$ and mass $M_M$:
(a) pion with $J=0$ and $M_\pi = 0.13957$ GeV, 
(b) $\rho$ meson with $J=1$ and $M_\rho = 0.7665$ GeV,
(c) $a_2$ meson with $J=2$ and $M_{a_2} = 1.3186$ GeV,
(d) $\omega_3$ meson with $J=3$ and $M_{\omega_3} = 1.67$ GeV~\cite{PDG22}.  

In Fig.~\ref{fig3} we present results for the massive bulk-to-boundary
propagator $\phi_{M_{J, \tau}}(Q^2+M_M^2,z)$ as a function of $Q^2$ and $J$
at fixed values of $z=1$ and $z=2$ GeV$^{-1}$, for leading twist $\tau = 2$.
Figure~\ref{fig3} shows that
increasing of $z$ leads to a more suppressed behavior of
the $\phi_{M_{J, \tau}}(Q^2+M_M^2,z)$, as expected. 

Next we check that the boson bulk-to-boundary
propagator with arbitrary $J$, dual to gauge bosons or
mesons, produces the correct power
scaling of hadronic form factors 
$F(Q^2) \sim 1/Q^{2 (\tau-1)}$ at large Euclidean values of $Q^2 = -q^2$.
In particular, we derive the master formulas for the transition
meson and baryon form factors $F_{V_JM_{\tau_1}M_{\tau_2}}(Q^2)$ 
and $F_{V_JB_{\tau_1}B_{\tau_2}}(Q^2)$, which are produced
by the integral over the holographic coordinate $z$
of the product of the
bulk-to-boundary propagator $V_J(Q^2+M^2,z)$ (off-shell
SM or NP bosons with quantum number $J$) or $\phi_{M_{J, \tau}}(Q^2+M_M^2,z)$
(off-shell meson with quantum numbers $J$ and $\tau$) 
and two hadron wave functions with arbitrary leading twists
$\tau_1$ and $\tau_2$: meson $\phi_{M_{\tau_1}}(z)$ and
$\phi_{M_{\tau_2}}(z)$, or baryon $\phi_{B_{\tau_1}}(z)$ and
$\phi_{B_{\tau_2}}(z)$ wave functions,
respectively~\cite{Brodsky:2007hb},\cite{Grigoryan:2007my}-\cite{Gutsche:2011vb}. 

First, we consider the case of a meson transition form factor induced
by off-shell SM or NP bosons, 
which in the soft-wall AdS/QCD model is given
by~\cite{Brodsky:2007hb,Grigoryan:2007my,Gutsche:2011vb}
\eq\label{formfactor_Jtau1tau2}
F_{V_JM_{\tau_1}M_{\tau_2}}(Q^2) &=& g_{V_JM_{\tau_1}M_{\tau_2}}  \, 
\int\limits_0^\infty dz \, V_J(Q^2+M^2,z) \, \phi_{M_{\tau_1}}(z) 
\, \phi_{M_{\tau_2}}(z) \,, 
\nonumber\\ 
&=& g_{V_JM_{\tau_1}M_{\tau_2}} \,
\frac{\Gamma\Big(J+\frac{\tau_1+\tau_2}{2}-1\Big)}{\Gamma(J)
  \, \sqrt{\Gamma(\tau_1-1) \, \Gamma(\tau_2-1)}}
 \, B\Big(a(M^2)+J,\frac{\tau_1+\tau_2}{2}-1\Big),
\en 
where $g_{V_JM_{\tau_1}M_{\tau_2}}$ is the normalization constant, fixed
by gauge invariance, from 
data or from phenomenological approaches. 
Equation~(\ref{formfactor_Jtau1tau2}) can be also used for the description of 
the coupling of an off-shell meson with total angular momentum $J$
with two mass-shell mesons with leading twists $\tau_1$ and $\tau_2$.
In the soft-wall AdS/QCD model,  the form factor~(\ref{formfactor_Jtau1tau2}) 
for the case $J=1$ was calculated for the first time 
in Ref.~\cite{Brodsky:2016uln}.

It can be seen that all the $Q^2$ 
dependence of the form factor $F_{V_JM_{\tau_1}M_{\tau_2}}(Q^2)$
is encoded in the beta function. Therefore, the large $Q^2$
behavior of $F_{V_JM_{\tau_1}M_{\tau_2}}(Q^2)$ is defined by
the corresponding behavior of the beta function. 
At large $Q^2 \gg M^2$ the form factor $F_{V_JM_{\tau_1}M_{\tau_2}}(Q^2)$
has the scaling independent on $J$, 
\eq\label{ff_largeQ2} 
F_{V_JM_{\tau_1}M_{\tau_2}}(Q^2) \sim
\frac{1}{Q^{2 \, (\frac{\tau_1+\tau_2}{2}-1)}}
\,.
\en
The $J$ dependence only remains in the coupling constant
$g_{V_JM_{\tau_1}M_{\tau_2}}$ and the factor
\eq
\frac{\Gamma\Big(J+\frac{\tau_1+\tau_2}{2}-1\Big)}{\Gamma(J)
    \, \sqrt{\Gamma(\tau_1-1) \, \Gamma(\tau_2-2)}} \,.
\en 
In the special case $\tau_1 = \tau_2 = \tau$ we reproduce
the result dictated by quark counting rules~\cite{Brodsky:1973kr}
\eq\label{ff_dia_largeQ2} 
F_{V_JM_{\tau}M_{\tau}}(Q^2) \sim \frac{1}{Q^{2 \, (\tau-1)}}
\,.
\en
Next, by analogy with the meson case,
we derive a baryon transition form factor induced
by off-shell SM or NP bosons and two on-shell
baryons. In the case of two on-shell baryons
one should take into account that the baryon AdS spinors are 
decomposed into two solutions: 
right-handed $\phi_{B_\tau}^{(r)}(z)$ and left-handed $\phi_{B_\tau}^{(\ell)}(z)$ 
chiral eigenstates~\cite{Abidin:2009hr,Vega:2010ns,Gutsche:2011vb,%
Gutsche:2012bp,Brodsky:2014yha}. Here, the leading twist of baryon field
$\tau$ is related to the angular orbital momentum $L$ as 
$\tau = 3 + L$, E.g., for $L=0$ baryons (e.g.,
nucleons with $J^P = \frac{1}{2}^+$ and $\Delta$ isobars
with $J^P = \frac{3}{2}^+$) the leading twist equals to $\tau = 3$.
For fixed internal spin $S$ the total angular momentum $J$ runs
from $|L-S|$ to $|L+S|$.
Therefore, by changing the value of $L$
we can generate the solutions for baryonic wave functions with any
required value of total angular momentum-parity $J^P$. 
In the soft-wall AdS/QCD approach the baryon wave functions
with specific leading twist have definite relations with
the corresponding meson wave functions.
In particular, the baryon wave function $\phi_{B_\tau}^{(r)}(z)$ 
coincides with the meson wave function $\phi_{M_\tau}(z)$, while
$\phi_{B_\tau}^{(l)}(z)$ is related to $\phi_{B_\tau}^{(r)}(z)$ as
$\phi_{B_\tau}^{(l)}(z) = \phi_{B_{\tau+1}}^{(r)}(z)$.
In particular~\cite{Gutsche:2011vb}, 
\eq\label{Btaunwf}
\phi_{B_\tau}^{(r)}(z)  &\equiv&
\phi_{M_\tau}(z) = \sqrt{\frac{2}{\Gamma(\tau-1)}}
\, \kappa^{\tau-1} \, z^{\tau-3/2} \, e^{-\kappa^2z^2/2} \,,
\nonumber\\
\phi_{B_\tau}^{(l)}(z)  &\equiv& \phi_{B_\tau+1}^{(r)}(z) =
\sqrt{\frac{2}{\Gamma(\tau)}}
\, \kappa^{\tau}   \, z^{\tau-1/2} \, e^{-\kappa^2z^2/2} 
\,. 
\en

The form factors describing the coupling of an external boson field
with total angular momentum $J$ and two baryons with leading
twists $\tau_1$ and $\tau_2$ and specific handedness $(r)$ or $(l)$
are calculated~\cite{Abidin:2009hr,Gutsche:2011vb} by analogy
with the case of meson form factors~(\ref{formfactor_Jtau1tau2}).
Then we get the following expression 
using this analogy, with the resulting formula looking very similar: 
\eq\label{VJBB}
F_{V_JB_{\tau_1}^{(i_1)}B_{\tau_2}^{(i_2)}}(Q^2) =
g_{V_JB_{\tau_1}^{(i_1)}B_{\tau_2}^{(i_2)}} \, 
\int\limits_0^\infty dz \, V_J(Q^2+M^2,z)
\, \phi_{B_{\tau_1}}^{(i_1)}(z) \, \phi_{B_{\tau_2}}^{(i_2)}(z) \,,
\en
where $g_{V_JB_{\tau_1}^{(i_1)}B_{\tau_2}^{(i_2)}}$ and $i_1, i_2 = l, r$
are the normalization constants, which are introduced
by analogy with the case of the meson form factors.

By analogy with the meson case we get
for $F_{V_JB_{\tau_1}^{(i_1)}B_{\tau_2}^{(i_2)}}(Q^2)$:
\eq\label{VJBB2} 
F_{V_JB_{\tau_1}^{(r)}B_{\tau_2}^{(r)}}(Q^2) &=&
g_{V_JB_{\tau_1}^{(r)}B_{\tau_2}^{(r)}} \, 
\frac{\Gamma\Big(J+\frac{\tau_1+\tau_2}{2}-1\Big)}{\Gamma(J)
\, \sqrt{\Gamma(\tau_1-1) \, \Gamma(\tau_2-1)}}
\, B\Big(a(M^2)+J,\frac{\tau_1+\tau_2}{2}-1\Big)
\,, \nonumber\\
F_{V_JB_{\tau_1}^{(r)}B_{\tau_2}^{(l)}}(Q^2) &=&
g_{V_JB_{\tau_1}^{(r)}B_{\tau_2}^{(l)}} \, 
\frac{\Gamma\Big(J+\frac{\tau_1+\tau_2-1}{2}\Big)}{\Gamma(J)
\, \sqrt{\Gamma(\tau_1-1) \, \Gamma(\tau_2)}}
\, B\Big(a(M^2)+J,\frac{\tau_1+\tau_2-1}{2}\Big)
\,, \nonumber\\
F_{V_JB_{\tau_1}^{(l)}B_{\tau_2}^{(r)}}(Q^2) &=&
g_{V_JB_{\tau_1}^{(l)}B_{\tau_2}^{(r)}} \, 
\frac{\Gamma\Big(J+\frac{\tau_1+\tau_2-1}{2}\Big)}{\Gamma(J)
\, \sqrt{\Gamma(\tau_1) \, \Gamma(\tau_2-1)}}
\, B\Big(a(M^2)+J,\frac{\tau_1+\tau_2-1}{2}\Big)
\,, \nonumber\\
F_{V_JB_{\tau_1}^{(l)}B_{\tau_2}^{(l)}}(Q^2) &=&
g_{V_JB_{\tau_1}^{(l)}B_{\tau_2}^{(l)}} \, 
\frac{\Gamma\Big(J+\frac{\tau_1+\tau_2}{2}\Big)}{\Gamma(J)
\, \sqrt{\Gamma(\tau_1) \, \Gamma(\tau_2)}}
\, B\Big(a(M^2)+J,\frac{\tau_1+\tau_2}{2}\Big)
\,. 
\en

At large $Q^2$, the form factors $F_{V_JB_{\tau_1}^{(i_1)}B_{\tau_2}^{(i_2)}}(Q^2)$
scale as
\eq\label{ff_largeQ2_baryon} 
F_{V_JB_{\tau_1}^{(r)}B_{\tau_2}^{(r)}}(Q^2) &\sim&
\frac{1}{Q^{2 \, (\frac{\tau_1+\tau_2}{2}-1)}}
\,, \nonumber\\
F_{V_JB_{\tau_1}^{(r)}B_{\tau_2}^{(l)}}(Q^2) &\sim&
F_{V_JB_{\tau_1}^{(l)}B_{\tau_2}^{(r)}}(Q^2) \, \sim \, 
\frac{1}{Q^{2 \, (\frac{\tau_1+\tau_2-1}{2})}}
\,, \nonumber\\
F_{V_JB_{\tau_1}^{(l)}B_{\tau_2}^{(l)}}(Q^2) &\sim&
\frac{1}{Q^{2 \, (\frac{\tau_1+\tau_2}{2})}}
\,.
\en 
Note that the left-handed baryon wave function
produces an extra $1/\sqrt{Q^2}$ falloff. 
In the limiting case $\tau_1 = \tau_2 = \tau$  we reproduce
the result dictated by the quark counting rules~\cite{Brodsky:1973kr} 
for the $F_{V_JB_{\tau_1}^{(r)}B_{\tau_2}^{(r)}}(Q^2)$ form factor
\eq\label{ff_dia_largeQ2_baryon} 
F_{V_JB_{\tau}^{(r)}B_{\tau}^{(r)}}(Q^2) \sim \frac{1}{Q^{2 \, (\tau-1)}}
\,.
\en
The other three form factors have extra
$1/\sqrt{Q^2}$ and $1/Q^2$ falloff, respectively  
\eq\label{ff_dia_largeQ2_baryon2} 
F_{V_JB_{\tau}^{(l)}B_{\tau}^{(r)}}(Q^2) \sim
F_{V_JB_{\tau}^{(r)}B_{\tau}^{(l)}}(Q^2) \sim
\frac{1}{Q^{2 \, (\tau-1/2)}}
\,, \qquad 
F_{V_JB_{\tau}^{(l)}B_{\tau}^{(l)}}(Q^2) \sim
\frac{1}{Q^{2 \tau}} 
\,. 
\en

Next we derive analytical expressions for the form factors describing
the direct coupling of three hadrons induced by an off-shell meson
with quantum numbers $J$ and $\tau$ and 
two on-shell hadrons (two mesons or two baryons) with
twists $\tau_1$ and $\tau_2$. As we pointed out 
before, the off-shell meson is described by the
bulk-to-boundary propagator $\phi_{M_{J, \tau}}(Q^2+M_M^2,z)$, while 
on-shell hadrons by the corresponding hadronic wave functions
with leading twists $\tau_1$ and $\tau_2$ defined before
in Eqs.~(\ref{Htaunwf}) and~(\ref{Btaunwf}).  

The coupling of off-shell meson $J \ge 1$ with two on-shell mesons reads 
\eq\label{FMMM}  
F_{M_{J, \tau}M_{\tau_1}M_{\tau_2}}(Q^2) &=&
g_{M_{J, \tau}M_{\tau_1}M_{\tau_2}} \,
\frac{\Gamma\Big(\frac{\tau+\tau_1+\tau_2}{2}+J-2\Big)}{\Gamma(J)
  \, \sqrt{\Gamma(\tau-1) \, \Gamma(\tau_1-1) \, \Gamma(\tau_2-1)}}
\nonumber\\
&\times& \int\limits_0^1  dx \, x^{a(M^2_M)+J-1} \, (1-x)^{\frac{\tau+\tau_1+\tau_2}{2}-3}
\, \Big(\frac{2}{3-x}\Big)^{\frac{\tau+\tau_1+\tau_2}{2}+J-3} \,,
\en
where $g_{M_{J, \tau}M_{\tau_1}M_{\tau_2}}$ is the normalization constant. 
For $J=0$ we get
\eq\label{FMMM0}  
F_{M_{0, \tau}M_{\tau_1}M_{\tau_2}}(Q^2) &=&
g_{M_{0, \tau}M_{\tau_1}M_{\tau_2}} \,
\frac{\Gamma\Big(\frac{\tau+\tau_1+\tau_2}{2}-1\Big)}
{\sqrt{\Gamma(\tau-1) \, \Gamma(\tau_1-1) \, \Gamma(\tau_2-1)}}
\nonumber\\
&\times& \int\limits_0^1  dx \, x^{a(M^2_M)} \, (1-x)^{\frac{\tau+\tau_1+\tau_2}{2}-3}
\, \Big(\frac{2}{3-x}\Big)^{\frac{\tau+\tau_1+\tau_2}{2}+J-2} \,. 
\en
At $Q^2 \to \infty$ the form factor~(\ref{FMMM})
is independent on the total angular momentum $J$ of off-shell meson
and scales as 
\eq\label{asymptotics_FMMM}
F_{M_{J, \tau}M_{\tau_1}M_{\tau_2}}(Q^2) \sim
\frac{1}{(Q^2)^{\frac{\tau+\tau_1+\tau_2}{2}-2}}
\,.
\en
The couplings of off-shell mesons with two on-shell baryons are
calculated by analogy with the case of the three-meson coupling
discussed above. We get the following relations
between three-meson and meson-two-baryon form factors:
\eq
F_{M_{J, \tau}B_{\tau_1}^{(r)}B_{\tau_2}^{(r)}}(Q^2) &=&
\frac{g_{M_{J, \tau}B_{\tau_1}^{(r)}B_{\tau_2}^{(r)}}}{g_{M_{J, \tau}M_{\tau_1}M_{\tau_2}}}
\, F_{M_{J, \tau}M_{\tau_1}M_{\tau_2}}(Q^2)
\,, \\
F_{M_{J, \tau}B_{\tau_1}^{(r)}B_{\tau_2}^{(l)}}(Q^2) &=&
\frac{g_{M_{J, \tau}B_{\tau_1}^{(r)}B_{\tau_2}^{(l)}}}{g_{M_{J, \tau}M_{\tau_1}M_{\tau_2+1}}}
\, F_{M_{J, \tau}M_{\tau_1}M_{\tau_2+1}}(Q^2)
\,, \\
F_{M_{J, \tau}B_{\tau_1}^{(l)}B_{\tau_2}^{(r)}}(Q^2) &=&
\frac{g_{M_{J, \tau}B_{\tau_1}^{(l)}B_{\tau_2}^{(r)}}}{g_{M_{J, \tau}M_{\tau_1+1}M_{\tau_2}}}
\, F_{M_{J, \tau}M_{\tau_1+1}M_{\tau_2}}(Q^2)
\,, \\
F_{M_{J, \tau}B_{\tau_1}^{(l)}B_{\tau_2}^{(l)}}(Q^2) &=&
\frac{g_{M_{J, \tau}B_{\tau_1}^{(l)}B_{\tau_2}^{(l)}}}{g_{M_{J, \tau}M_{\tau_1+1}M_{\tau_2+1}}}
\, F_{M_{J, \tau}M_{\tau_1+1}M_{\tau_2+1}}(Q^2)
\,.
\en

\subsection{Fermion bulk-to-boundary propagator}

Next we derive the fermion bulk-to-boundary propagator, e.g. the
corresponding off-shell baryons. As in the bosons case,
first we derive the massless propagator and then
extend it to the finite mass case by analogy with bosons. 
The soft-wall AdS/QCD action 
has been derived in Ref.~\cite{Gutsche:2011vb} ,
for fermions with higher $J \ge 5/2$: 
\eq
\hspace*{-.75cm}
S_{J} &=&  \int d^dx dz \, \sqrt{g} \, e^{-\varphi(z)} \,
\biggl[ \frac{i}{2} \bar\Psi^{N_1 \cdots N_{J-1/2}}(x,z)
\epsilon_a^M \Gamma^a {\cal D}_M \Psi_{N_1 \cdots N_{J-1/2}}(x,z)
\nonumber\\
&-& \frac{i}{2}
({\cal D}_M\Psi^{N_1 \cdots N_{J-1/2}}(x,z))^\dagger
\Gamma^0 \epsilon_a^M \Gamma^a \Psi_{N_1 \cdots N_{J-1/2}}(x,z)
- \bar\Psi^{ N_1 \cdots N_{J-1/2}}(x,z) \Big(\mu + V_F(z)\Big)
\Psi_{N_1 \cdots N_{J-1/2}}(x,z) \biggr] \,,
\en
where $\Psi_{N_1 \cdots N_{J-1/2}}$ is the spin-tensor field,
$g = |{\rm det} g_{MN}| = e^{10 A(z)}$,
$\mu = (L+3/2)/R$ is the bulk fermion mass,
$V_F(z) = \varphi(z)/R$ is the dilaton field-dependent effective
potential, and ${\cal D}_M$ is the covariant derivative acting on
the spin-tensor field defined as 
\eq
{\cal D}_M \Psi_{N_1 \cdots N_{J-1/2}} &=&
\partial_M \Psi_{N_1 \cdots N_{J-1/2}}
- \Gamma^K_{MN_1} \Psi_{K N_2 \cdots N_{J-1/2}}
- \cdots
\nonumber\\
&-& \Gamma^K_{MN_{J-1/2}} \Psi_{N_1 \cdots N_{J-3/2} K}
- \frac{1}{8} \omega_M^{ab} [\Gamma_a, \Gamma_b]
\Psi_{N_1 \cdots N_{J-1/2}} \,.
\en
Here $\omega_M^{ab}$ and $\Gamma^K_{MN}$ are the spin and
affine connections, which are defined and related as
\eq
\omega_M^{ab} = A^\prime(z) \,
(\delta^a_z \delta^b_M - \delta^b_z \delta^a_M) 
= \epsilon_K^a \Big( \partial_M\epsilon^{Kb}
+ \epsilon^{Nb} \, \Gamma_{MN}^K \Big) \,, 
\en
$\Gamma^a = (\gamma^\mu,-i \gamma^5)$ and $\Gamma^0 = \gamma^0$
are the Dirac matrices.

Next, decomposing the fermion field in left- and right-chirality
components 
\eq
  \Psi_{\mu_1 \cdots \mu_{J-1/2}}(x,z) = 
  \Psi^{(l)}_{\mu_1 \cdots \mu_{J-1/2}}(x,z)
+ \Psi^{(r)}_{\mu_1 \cdots \mu_{J-1/2}}(x,z)
\,, \quad
\Psi^{(l/r)} = \frac{1 \mp \gamma^5}{2} \Psi \,, \quad
\gamma^5 \Psi^{(l/r)} = \mp \Psi^{(r/l)} 
\en
and performing the Fourier transformation for the
$\Psi^{(l)}(x,z)$ and $\Psi^{(r)}(x,z)$ fields in terms
of left- and right-handed bulk-to-boundary propagators
$F^{(l)}_L(-q^2,z)$ and $F^{(r)}_L(-q^2,z)$
with orbital momentum $L$ (lower index) and left $(l)$
and right $(r)$ chirality (superscript indices) 
\eq
\Psi^{(l/r)}_{\mu_1 \cdots \mu_{J-1/2}}(x,z)
= \int \frac{d^4 q}{(2\pi)^4}  \,
e^{-iqx} \, \Psi^{(l/r)}_{\mu_1 \cdots \mu_{J-1/2}}(q)  \,
F^{(l/r)}_L(-q^2,z), 
\en 
we derive the equations of motion for the massless 
fermion bulk-to-boundary propagators
$F^{(r)}_L(-q^2,z)$ and $F^{(l)}_L(-q^2,z)$
\eq\label{bulkF}
& &\partial_z \biggl( e^{-B^{(r)}(z)}  \, \partial_z \,
F_L^{(r)}(-q^2,z)\biggr)
+ e^{-B^{(r)}(z)}  \, q^2 \, F_L^{(r)}(-q^2,z) = 0 \,,
\nonumber\\
& &\partial_z \biggl( e^{-B^{(l)}(z)}  \, \partial_z \,
F_L^{(l)}(-q^2,z)\biggr)
+ e^{-B^{(l)}(z)}  \, q^2 \, F_L^{(l)}(-q^2,z) = 0 \,.
\en
Here
$B^{(r)}(z) = \phi(z) - (2 L + 1) A(z)$ and
$B^{(l)}(z) = \phi(z) - (2 L + 3) A(z)$. 
The equations of motion and their solutions for the fermion
propagators are similar to those
for the boson propagators~(\ref{bulkV}). 
We also get the same equations and solutions for
the fermion propagators with lower values of 
$J=\frac{1}{2}$ and $\frac{3}{2}$, whose actions
were already discussed in Ref.~\cite{Gutsche:2011vb}. 

We establish the following
relations between the solutions for massless boson and fermion
bulk-to-boundary propagators:
\eq\label{propagator_VFJ}
F_L^{(r)}(Q^2,z) = F_{L-1}^{(l)}(Q^2,z) = V_{L+1}(Q^2,z)
= \frac{1}{\Gamma(L+1)} \,
\int\limits_0^\infty  dy \, y^{L} \, e^{-y} \,
\biggl(\frac{y}{y+\kappa^2 z^2}\biggr)^{a} 
\,.
\en
As in the case of boson propagators, the fermion ones are
also properly normalized:
\eq
F_L^{(r)}(0,z) = F_L^{(l)}(0,z) =
F_L^{(r)}(Q^2,0) = F_L^{(l)}(Q^2,0) = 1 \,,
\en
and they also vanish at $z \to \infty$. By analogy with the boson case
we include the finite mass $M$ in the fermion bulk-to-boundary propagator,
via the extension $Q^2 \to Q^2 + M^2$. 

As application of the fermion bulk-to-boundary propagators,
we consider only the case of their duals --- off-shell baryons
with quantum numbers of total angular momentum $J$ and mass $M_B$.
In particular, we calculate the form factors describing 
the coupling of an off-shell baryon with a pair of
on-shell meson and baryon.
By analogy with the mesons case
we define the bulk-to-boundary propagator
dual to massive baryon with artbitrary $J$ as
the product of the fermion bulk-to-boundary propagator
and baryon wave function with specific handedness $i = l, r$: 
\eq\label{V_baryon}
\phi_{B_{L, \tau}}^{(i)}(Q^2+M_B^2,z) = F_L^{(i)}(Q^2+M_B^2,z)
\, \phi_{B_\tau}^{(i)}(z) \,. 
\en 
In this case we have four possibilities,
corresponding to the two possible handedness of the fermion
bulk-to-boundary propagator and the baryon: 
(i) right-handed off-shell baryon couples
with right-handed on-shell baryon,
(ii) right-handed off-shell baryon couples
with left-handed on-shell baryon,
(iii) left-handed off-shell baryon couples
with right-handed on-shell baryon,
(iv) left-handed off-shell baryon couples
with left-handed on-shell baryon. 
For these four possibilities
one can produce four types of form factors: 
\eq\label{FJMB}
F_{B_{L, \tau}^{(i_1)}M_{\tau_1}B_{\tau_2}^{(i_2)}}(Q^2) =
g_{B_{L, \tau}^{(i_1)}M_{\tau_1}B_{\tau_2}^{(i_2)}} \, 
\int\limits_0^\infty dz \, \phi_{B_{L, \tau}}^{(i)}(Q^2+M_B^2,z) 
\, \phi_{M_{\tau_1}}(z) \, \phi_{B_{\tau_2}}^{(i_2)}(z)
\,,
\en
where $g_{B_{L, \tau}^{(i_1)}M_{\tau_1}B_{\tau_2}^{(i_2)}}$
are the normalization constants. 

Baryon form factors $F_{B_{L, \tau}^{(i_1)}M_{\tau_1}B_{\tau_2}^{(i_2)}}(Q^2)$
are related to meson form factors
$F_{M_{J, \tau}M_{\tau_1}M_{\tau_2}}(Q^2)$~(\ref{FMMM}) as 
\eq\label{FJBB2}
F_{B_{L, \tau}^{(r)}M_{\tau_1}B_{\tau_2}^{(r)}}(Q^2) &=&
\frac{g_{B_{L, \tau}^{(r)}M_{\tau_1}B_{\tau_2}^{(r)}}}{g_{M_{L+1, \tau}M_{\tau_1}M_{\tau_2}}}
\, F_{M_{L+1, \tau}M_{\tau_1}M_{\tau_2}}(Q^2)
\,, \\
F_{B_{L, \tau}^{(r)}M_{\tau_1}B_{\tau_2}^{(l)}}(Q^2) &=&
\frac{g_{B_{L, \tau}^{(r)}M_{\tau_1}B_{\tau_2}^{(l)}}}{g_{M_{L+1, \tau}M_{\tau_1}M_{\tau_2+1}}}
\, F_{M_{L+1, \tau}M_{\tau_1}M_{\tau_2+1}}(Q^2)
\,, \\
F_{B_{L, \tau}^{(l)}M_{\tau_1}B_{\tau_2}^{(r)}}(Q^2) &=&
\frac{g_{B_{L, \tau}^{(l)}M_{\tau_1}B_{\tau_2}^{(r)}}}{g_{M_{L+2, \tau+1}M_{\tau_1}M_{\tau_2}}}
\, F_{M_{L+2, \tau+1}M_{\tau_1}M_{\tau_2}}(Q^2)
\,, \\
F_{B_{L, \tau}^{(l)}M_{\tau_1}B_{\tau_2}^{(l)}}(Q^2) &=&
\frac{g_{B_{L, \tau}^{(l)}M_{\tau_1}B_{\tau_2}^{(l)}}}{g_{M_{L+2, \tau+1}M_{\tau_1}M_{\tau_2+1}}}
\, F_{M_{L+2, \tau+1}M_{\tau_1}M_{\tau_2+1}}(Q^2)
\,.
\en
At large $Q^2$ these form factors scale as
\eq
F_{B_{L, \tau}^{(r)}M_{\tau_1}B_{\tau_2}^{(r)}}(Q^2) &\sim&
\frac{1}{Q^{2 \, (\frac{\tau+\tau_1+\tau_2}{2}-1)}}
\,, \nonumber\\
F_{B_{L, \tau}^{(l)}M_{\tau_1}B_{\tau_2}^{(r)}}(Q^2) &\sim&
F_{B_{L, \tau}^{(r)}M_{\tau_1}B_{\tau_2}^{(l)}}(Q^2) \, \sim \,
\frac{1}{Q^{2 \, (\frac{\tau+\tau_1+\tau_2-1}{2})}}
\,, \nonumber\\
F_{B_{L, \tau}^{(l)}M_{\tau_1}B_{\tau_2}^{(l)}}(Q^2) &\sim&
\frac{1}{Q^{2 \, (\frac{\tau+\tau_1+\tau_2}{2})}}
\,.
\en
For example, for the coupling with leading twist-2 meson and
leading twist-3 baryon, one gets:
\eq
F_{B_{L, 3}^{(r)}M_{2}B_{3}^{(r)}}(Q^2) &\sim&
\frac{1}{Q^6}
\,, \nonumber\\
F_{B_{L, 3}^{(l)}M_{2}B_{3}^{(r)}}(Q^2) &\sim&
F_{B_{L, 3}^{(r)}M_{2}B_{3}^{(l)}}(Q^2) \, \sim \, 
\frac{1}{Q^7}
\,, \nonumber\\
F_{B_{L, 3}^{(l)}M_{2}B_{3}^{(l)}}(Q^2) &\sim&
\frac{1}{Q^8}
\,.
\en
As was expected, the left-handed baryon wave function
produces an extra falloff $1/\sqrt{Q^2}$ in comparison with the
right-handed one. 

\section{Conclusion}

We proposed an extension of
thesoft-wall AdS/QCD model for the calculation of boson 
and fermion bulk-to-boundary propagators with arbitrary total angular
momentum $J$. 
Starting from AdS/QCD actions for boson and fermion fields with arbitrary $J$,
we derived EOMs for the massless boson and fermion bulk-to-boundary propagators.
Next we include finite masses of the bulk-to-boundary propagators,
by shifting the square of the momentum as $- q^2 = Q^2 \to - q^2 + M^2 = Q^2 + M^2$,
where $M$ is the mass of the SM or NP fields or hadrons. Bulk-to-boundary propagators
obey known and required properties of charge conservation, local limit, and confinement.

The bulk-to-boundary propagators are dual to off-shell SM (NP) fields or off-shell hadrons. 
This allows one to calculate form factors describing the coupling of two on-shell hadrons
(mesons or baryons) with an off-shell SM (NP) field or hadron. The produced form factors 
are consistent, at large $Q^2$, with the constituent counting rules~\cite{Brodsky:1973kr}.
In the case of the bulk-to-boundary propagators, dual to SM (NP) fields, the application
of our formalism is relevant for the values of $J=0,1,2$. For the case of
bulk-to-boundary propagators dual to off-shell hadrons, we are not limited
by upper values of $J$, because hadrons (both mesons and baryons)
with higher $J$
have been searched experimentally and predicted or studied in theoretical
approaches~\cite{PDG22}. According to the Particle Data Group~\cite{PDG22},
mesons up to $J=6$ and baryons up to $J=15/2$ are known.

We derived the set of analytical formulas describing hadronic form factors with one
off-shell and two on-shell particles.
This lead to a unique opportunity to study the off-shell behavior
of hadronic form factors. Therefore it provides useful insight to lattice QCD
and effective field theories, where direct couplings of hadrons are calculated
from the first principles or provide input parameters for phenomenological Lagrangians.
Our formalism can be straightforwardly extended for study of hadronic form factors
with two and three off-shell particles. 

\begin{acknowledgments}

This work was funded by BMBF (Germany)
``Verbundprojekt 05P2021 (ErUM-FSP T01) -
Run 3 von ALICE am LHC: Perturbative Berechnungen
von Wirkungsquerschnitten f\"ur ALICE''
(F\"orderkennzeichen: 05P21VTCAA),
by ANID PIA/APOYO AFB180002 and AFB220004 (Chile),
by FONDECYT (Chile) under Grants No. 1191103, 
No. 1230160 and No. 1230391
and by ANID$-$Millen\-nium Program$-$ICN2019\_044 (Chile).

\end{acknowledgments}

\end{document}